# A Paradox of Whistling Atmospherics

A.V. Guglielmi[1], B.I. Klain[2], A.S. Potapov[3]

[1] Schmidt Institute of Physics of the Earth, Russian Academy of Sciences, Moscow, Russia, guglielmi@mail.ru
[2] Borok Geophysical Observatory, the Branch of Schmidt Institute of Physics of the Earth, Russian Academy of Sciences, Borok, Yaroslavl Region, Russia, klb314@mail.ru
[3] Institute of Solar-Terrestrial Physics of Siberian Branch of Russian Academy of Sciences, Irkutsk, Russia, potapov@iszf.irk.ru

**Abstract**

Analyzing paradoxes is interesting and instructive. Sometimes the analysis leads to non-trivial results. This methodological note sets out a paradox arising in the theory of propagation of electromagnetic waves in moving plasmas. The paradox is interesting in itself, and, generally speaking, it should be taken into account when analyzing geoelectromagnetic waves. The paradox is as follows: contrary to expectations, the group velocity of the waves is the same in the comoving and laboratory frames of reference. The condition for the appearance of the paradox is the quadratic dependence of the frequency on the wave number. A paradoxical property manifests itself in the theory of the propagation of radio waves (in particular, whistling atmospherics), Langmuir waves and Alfvén waves. From a cognitive point of view, it is interesting that the paradox can be traced in relation to de Broglie waves. An explanation of the paradox is proposed.

## INTRODUCTION

The fundamental concept of the group velocity of wave propagation

$$\mathbf{v}_g = \frac{d\omega}{d\mathbf{k}} \qquad (1)$$

was introduced by Rayleigh and Stokes independently in the 70s of the 19th century. Here $\mathbf{k}$ is the wave vector, $\omega$ is the frequency associated with $\mathbf{k}$ by the dispersion relation $\omega = \omega(\mathbf{k})$. The projection of the group velocity of electromagnetic waves on the direction of propagation can be expressed through the group refractive index $n_g$:

$$v_g = \frac{c}{n_g} \qquad (2)$$

Here, $n_g = \partial \omega n / \partial \omega$, $c$ is the speed of light [Landau, Lifshits, 1982]. The usual refractive index $n(\omega, \theta)$ in an anisotropic medium depends on the angle $\theta$ between the vector **k** and the axis of anisotropy, which in magnetoactive plasma coincides with the tangent to the field line of external magnetic field **B** [Ginzburg, 1960]. In general, the direction of group velocity does not coincide with the direction of **k** vector.

Let us introduce a comoving frame of reference, in which the medium (plasma) is motionless, and a laboratory frame of reference, moving with nonrelativistic velocity **u**. By the Doppler formula

$$\omega = \mathbf{ku} + \omega' \qquad (3)$$

where $\omega'$ is the frequency in the laboratory frame of reference. In accordance with (1) and (3), the group velocity is transformed according to the law

$$\mathbf{v}'_g = \mathbf{v}_g - \mathbf{u} \qquad (4)$$

when passing from a comoving system to a laboratory one.

The paradox, which will be discussed in this methodological note, is interesting in itself, and, generally speaking, it should be borne in mind when analyzing geoelectromagnetic waves. The paradox was first discovered by analyzing whistling atmospherics [Guglielmi, 1963]. Here we will find the condition for the paradox appearance and show that the paradox also arises in the theory of propagation of some other types of waves.

The essence of the paradox is as follows. Under the condition that we indicate below, the group velocity does not change when passing from a comoving frame to a moving laboratory frame of reference:

$$\mathbf{v}'_g = \mathbf{v}_g \qquad (5)$$

This statement, it would seem, is sharply contrary to common sense. But it turns out that our statement in some cases acquires a meaningful value.

## CONDITION FOR THE PARADOX APPEARANCE

Let us find the form of the dispersion relation at which the paradoxical property of the group velocity manifests itself. For this, we use the series expansion of the refractive index in the laboratory frame of reference

$$n'(\omega') = n(\omega) - \left.\frac{\partial \ln n'}{\partial \ln \omega'}\right|_\omega \frac{u}{c} \qquad (6)$$

Here $u$ is the projection of the observer's velocity vector onto the direction of wave propagation. The first term in the right-hand side of (6) is not primed for the reason that $\mathbf{k}' = \mathbf{k}$ at nonrelativistic velocity of motion. We limited ourselves to two terms of the series under the assumption that the observer's velocity is much less than the phase velocity of the wave, i.e. $u \ll c/n$.

According to the inverse function theorem, the dispersion equation can be represented in the form $k = f(\omega)$. Taking into account (6), the condition $v'_g = v_g$ leads to a nonlinear differential equation for the function $f(\omega)$

$$f \frac{d^2 f}{d\omega^2} + \left(\frac{df}{d\omega}\right)^2 = 0 \qquad (7)$$

It has an analytical solution which gives us

$$\omega = a k^2 + b \qquad (8)$$

Here $a$ and $b$ are arbitrary constants, and $a \neq 0$. Thus, the equality $v'_g = v_g$ is fulfilled provided that the frequency quadratically depends on the wave number.

## WHISTLING ATMOSPHERICS

A whistling atmospheric is excited by a lightning discharge in the troposphere and, propagating in the magnetosphere along the geomagnetic field line, crosses the equatorial plane and reaches a conjugate point on the Earth's surface in the other hemisphere [Likhter et al., 1988]. As it propagates, the impulse signal spreads out in such a way that at the output of the radio receiver we hear a whistle of a falling tone.

In the approximation of geometrical optics, the basic laws governing the propagation of a whistling atmospheric are derived using the Storey formula

$$n = \frac{\omega_0}{\sqrt{\omega \Omega \cos\theta}} \qquad (9)$$

Here $\omega_{0e}$ is the plasma Langmuir frequency, $\Omega$ is the electron gyrofrequency, $\omega \ll \Omega$, $n^2 \gg 1$ [Gershman, Ugarov, 1960]. Let us assume that the plasma moves along the field lines of the external magnetic field $\mathbf{B}$. This is a fairly realistic situation, since there is a flow of plasma along the geomagnetic field lines from the summer hemisphere to the winter one. From formula (9) at $\theta = 0$ it follows that

$$\omega = \alpha k^2 \qquad (10)$$

where $\alpha = \Omega k_0^{-2}$, $k_0 = \omega_0 / c$, which coincides with (8), and $a = \alpha$ and $b = 0$. Thus, the paradox $v'_g = v_g$ arises since the dispersion relation for whistling atmospherics has the form of a quadratic dependence of the frequency on the wave number.

Lightning discharges excite not only whistling atmospherics, but also so-called spherics [Guglielmi, Pokhotelov, 1996]. In contrast to the whistling atmospheric, the spheric propagates not in the magnetosphere, but in the Earth-ionosphere waveguide. At frequencies close to the critical frequency of the waveguide, condition (8) is satisfied, i.e. the frequency quadratically depends on the wave number, and thus we are again faced with a paradox.

## EXAMPLES OF WAVES WITH SQUARE LAW OF DISPERSION

*De Broglie waves*. An interesting example of waves with dispersion relation (8) is provided by quantum mechanics. From the Schrödinger equation

$$i\hbar \frac{\partial \psi}{\partial t} = -\frac{\hbar^2}{2m} \nabla^2 \psi + V\psi \tag{11}$$

it follows that the dispersion law for a free electron has the form

$$\omega = \frac{\hbar}{2m} k^2, \tag{12}$$

where $m$ is the electron mass, $\hbar$ is the Planck constant [Landau, Lifshits, 1989]. Formally, this coincides with the dispersion relation (10) for whistlers, so that $v'_g = v_g$.

*Langmuir waves*. In a collisionless plasma, the longitudinal Langmuir waves experience Landau damping [Pitaevsky, Lifshits, 1979]. We will neglect damping. Then the dispersion relation will have the form

$$\omega = \omega_0 \left(1 + \frac{3}{2} D^2 k^2 \right). \tag{13}$$

Here $D$ is the Debye radius [Kadomtsev, 1968]. Thus, Langmuir waves can be added to our collection of waves with a quadratic dispersion law.

*Transverse electromagnetic waves*. In an isotropic plasma, the square of the refractive index of transverse electromagnetic waves is

$$n^2 = 1 - \frac{\omega_0^2}{\omega^2}. \tag{14}$$

Accordingly, the dispersion equation has the form

$$\omega = \sqrt{\omega_0^2 + c^2 k^2}. \tag{15}$$

The plasma transparency region is limited from below by the cutoff frequency $\omega_0$. The group velocity in the comoving frame is

$$v_g = c\sqrt{1 - \frac{\omega_0^2}{\omega^2}}. \qquad (16)$$

When approaching from above to the boundary of the transparency region ($\omega \to \omega_0$), the group velocity tends to zero ($v_g \to 0$), so that any small motion of the medium relative to the observer will radically change the group delay time of the wave. But it is precisely near the boundary of the transparency region that the frequency quadratically depends on the wave number

$$\omega = \omega_0 + \frac{c^2}{2\omega_0}k^2, \qquad (17)$$

so our paradox is fully manifested in this case.

*Alfvén waves*. Let us show that the dispersion relation (8) is fulfilled in the near-Earth plasma not only for whistling atmospherics, but under certain conditions it is also fulfilled for Alfvén waves. It would seem that this is impossible, since, as is known from magnetohydrodynamics, the dispersion law has the form

$$\omega = c_A k_\parallel, \qquad (18)$$

which has nothing to do with (8). Here $c_A$ is the Alfvén velocity, $k_\parallel = (\mathbf{kB})/B$ is the longitudinal component of the wave vector [Alfvén, 1950]. But we will go through two procedures. First, we introduce two reflecting surfaces orthogonal to the field lines of the external magnetic field. As a result, the space between the surfaces forms a resonator with a discrete spectrum of oscillations $\omega_i = \pi i c_A / l$, where $i = 1, 2, 3, ...$, $l$ is the distance between the reflection points. We obtained in a simplified form the spectrum of oscillations of the ionospheric MHD resonator [Potapov et al., 2021]. Second, let us take into account the weak transverse dispersion of Alfvén waves [Guglielmi, Pokhotelov, 1996]. Then, in the cold-plasma approximation, instead of (18), we have

$$\omega_i(k_\perp) = \frac{\pi i c_A}{l}\left[1 - \left(\frac{k_\perp}{k_0}\right)^2\right]. \qquad (19)$$

Here $k_\perp$ is the value of the transverse component of the wave vector, $k_\perp^2 \ll k_0^2$. We see that the function $\omega_i(k_\perp)$ quadratically depends on $k_\perp$. Therefore, in accordance with (8) $v'_{g\perp} = v_{g\perp}$.

Note that the quadratic term enters into dispersion relation (19) with a negative sign. From this it is not difficult to deduce the consequence: $\mathbf{\kappa}_\perp \mathbf{v}_{g\perp} < 0$, i.e. 2D vectors $\mathbf{\kappa}_\perp$ and $\mathbf{v}_{g\perp}$ are antiparallel to each other. This property of Alfvén waves is characteristic of cold plasma. At the periphery of the magnetosphere, the plasma is hot, but the quadratic nature of the transverse dispersion remains. However, in this case $\mathbf{\kappa}_\perp \mathbf{v}_{g\perp} > 0$ and, accordingly, the vectors $\mathbf{\kappa}_\perp$ and $\mathbf{v}_{g\perp}$ are parallel to each other. Equality $v'_{g\perp} = v_{g\perp}$ is also satisfied in this case due to the quadratic nature of the dispersion law. The case of hot plasma is of certain interest from the point of view of the theory of a specific ion-cyclotron resonator, presumably existing in the equatorial vicinity of the radiation belt [Guglielmi et al., 2000; Guglielmi, Potapov, 2012, 2021].

## DISCUSSION

Analyzing paradoxes is interesting and instructive. The discovery of paradoxes stimulates a deeper understanding of the theory, and sometimes leads to new results. For example, Shezo's photometric paradox and the Neumann-Seliger gravitational paradox partly contributed to obtaining nontrivial results in cosmophysics.

We found a paradox by doing thought experiments with waves whose frequency is quadratically dependent on the wave number. It is not entirely clear whether the paradox is essential in real experiments. In this sense, our note has a methodological character.

Apparently, the easiest way would be to demonstrate the paradox in laboratory experiments on the excitation and propagation of Langmuir waves. However, the scientific purpose of such experiments is not clear. In natural experiments, the paradox can probably also manifest itself, but the question must be studied separately in each specific case, paying special attention to the physical side of the matter. For example, it may be necessary to take into account the anabatic or katabatic wind in the vicinity of the point of reflection of radio waves from the ionosphere when interpreting the results of ionospheric sounding.

Let us consider in more detail the MHD ionospheric resonator, known in the literature as IAR (ionospheric Alfvén resonator). (For the structure and properties of the resonator, see the review [Potapov et al., 2021]). A beam of waves originating in the outer radiation belt [Guglielmi, Potapov, 2021] or in the interplanetary medium ahead of the magnetosphere front [Guglielmi, Potapov, 2012] and incident on the ionosphere excites Alfvén and magnetosonic waves in the IAR. It seems to us that in the vicinity of the beam intrusion point, the ionospheric resonator acts on magnetosonic waves in approximately the same way as a Fabry-Perrot resonator acts on light waves. Relation (8) is approximately satisfied and, in principle, the paradox of interest to us can be observed.

One nuance should not be overlooked when considering resonances in IAR. If $k_\perp = 0$, then the upper wall of the resonator reflects magnetosonic waves in quite the same way as it happens with Alfvén waves. However, at a nonzero and sufficiently large $k_\perp$ value, the wave turns in the opposite direction not due to reflection from a region with a reduced plasma density, but due to refraction. This radically changes the spectral structure of the resonator: for sufficiently large $k_\perp$ values, the ionosphere should in fact be considered as a waveguide channeling magnetosonic waves along the earth's surface [Nishida, 1980].

It would seem that what was said above can be repeated in relation to Alfvén waves excited in the IAR. Formally, this is true, and the question is of certain theoretical interest, but in fact it is impossible to observe the horizontal propagation of Alfvén waves. The point is that the deviation of the beam, i.e. its deviation from the geomagnetic field line on the way from the lower wall of the IAR to the upper wall and back is so small that it is practically impossible to register the effect of horizontal propagation. Two irresistible circumstances prevent this from happening. It follows from (19) that $v_g$ depends bilinearly on $i$ and $k_\perp$. However, with an increase in the harmonic number $i$, the upper wall of the IAR loses its reflectivity and the Q-factor of the resonator rapidly decreases with $i$ increasing. As for the possibility of increasing $|v_g|$ by $k_\perp$ increasing, it is fundamentally limited by the condition $k_\perp \delta < 1$, where $\delta$ is the thickness of the air gap between the ground and the ionosphere. If the condition is violated, it is impossible to observe Alfvén oscillations from the ground. Inside the ionosphere, there is no specified limitation, and we can expect the existence of a very fine transverse structure of resonances, which is not visible from the ground.

Another interesting feature of Alfvén waves is the following: Alfvén caustics do not touch anywhere on surfaces orthogonal to the geomagnetic field lines [Guglielmi, 1989]. Hence, it follows that the reflection of waves from the upper wall of the resonator due to refraction is impossible for any value of $k_\perp$. Reflection occurs only as a result of violation of the conditions of applicability of geometric optics. This property of irreversibility, or better to say "non-reversal" of Alfvén rays, turns out to be very strong. The property is unique, although we know that it is lost when Alfvén waves propagate in a solid. Indeed, the dispersion law, similar to law (18), in a solid takes the form

$$\omega = \sqrt{c_A^2 k_\parallel^2 + c_t^2 k^2} \ . \tag{20}$$

Here $c_t$ is the velocity of transverse elastic waves. It can be seen that, from a theoretical point of view, in this case there are caustics orthogonal to the field lines of external magnetic field. However, according to our estimates, $c_t$ exceeds $c_A$ by many orders of magnitude in all known solids in the Universe, including in the pulsar crust. This means that the magnetic field has practically no effect on the propagation of transverse elastic waves.

Paradoxes often arise in the study of wave phenomena. At the end of this section, let us point out two more paradoxes that are quite worthy of careful study. It is shown in [Guglielmi, Zotov, 2020] that the Joule heating of a conducting half-space simulating the Earth's crust by an electromagnetic wave of cosmic origin does not depend on the electrical conductivity of rocks. The second paradox arises when analyzing the propagation of a wave pulse in a nonequilibrium (amplifying) medium. It turns out that the leading edge of a pulse can propagate at an arbitrarily high speed, even at a superluminal velocity (see the monograph [Guglielmi, 1979] and the literature cited therein).

## CONCLUSION

We have considered an interesting paradox arising in the theory of electromagnetic wave propagation. It gave us the opportunity to freely discuss a number of issues in the theory of propagation. The following explanation of the paradox seems to us reasonable. In essence, both formulas $v'_g = v_g - u$ and $v'_g = v_g$ are valid, each in its own area of applicability. If the velocity of a narrowband wave packet is measured with a broadband receiving equipment, then it is quite obvious that the velocity $v'_g = v_g - u$ will be measured. Velocity $v'_g = v_g$ will be measured when a broadband signal propagates in a square dispersion medium and the receivers are selective, tuned to a fixed frequency.

We express our sincere gratitude to O.D. Zotov and F.Z. Feygin for their interest in the study and stimulating discussions. This work was carried out with financial support from the Russian Foundation for Basic Research in the framework of project No. 19-05-00574, as well as from state assignments programs of the IPE RAS and ISTP SB RAS.